\documentclass[conference,10pt]{IEEEtran}
\usepackage[T1]{fontenc}
\usepackage{graphicx}
\usepackage{times}
\usepackage{cite}
\usepackage{amsmath,amssymb}
\usepackage{bm}
\usepackage{subfig}
\usepackage{enumerate}

\IEEEoverridecommandlockouts
\begin{document}
\newtheorem{theorem}{Theorem}
\newtheorem{lemma}[theorem]{Lemma}
\newtheorem{mydef}{Definition}

\title{Mitigating timing errors in time-interleaved ADCs: a signal conditioning approach}

\author{\authorblockN{Abhishek Ghosh, \textit {Student Member, IEEE} and Sudhakar Pamarti, \textit {Member, IEEE}}
\authorblockA{Department of Electrical Engineering, University of California, Los Angeles, CA 90095\\
Email:\{abhishek,spamarti\}@ee.ucla.edu}  \vspace*{-0.2in}}

\maketitle



\thispagestyle{empty}


%

\begin{abstract}
\par Novel techniques based on signal-conditioning are presented to mitigate timing errors in time-interleaved ADCs. A theoretical bound on the achievable spurious signal content, on applying the techniques, is also derived. Behavioral simulations corroborating the same are presented.
\end{abstract}
\section{Introduction}
\par Analog-to-digital converters (ADCs) form the interface between natural systems, which are essentially analog, and the digital regime of processing machines. Ever-escalating data rates have imposed severe challenges on the design of ADCs in terms of bandwidths of the signals that need to be digitized catering to communication circuits. Furthermore, complex modulation schemes invoked for higher spectral efficiency engender signals having very large dynamic ranges \cite{Rappaport}. Such signals demand to be digitized with a very high resolution in order to preserve their fidelity. Consequently, modern ADCs need to be of very high resolution (>12 bits) operating at extremely high speeds (in the GHz) to cater to the evolving communication sector.
\par To that end, several architectures have been proposed which can operate on signals having large bandwidths as well as high dynamic ranges. Time-interleaved architectures have proved to be one of the most suitable candidates for this purpose. Let's take a closer look at a time-interleaved ADC.

\subsection{Time-Interleaved ADCs: modeling}
\par The basic architecture is illustrated in Fig. 1. The incoming signal $x(t)$ is processed by $L$ parallel branches, such that each branch operates on the signal with a time-period $LT_S$ where $T_S$ is the overall sampling time for the ADC, satisfying $F_S=1/T_S$ ($F_S$ is the sampling frequency). It is assumed that $x(t)$ is a band-limited signal with a bandwidth $B$ such that $F_S > 2B$ (the familiar Nyquist relation). The main advantage of the time-interleaved architecture hence becomes evident. Since each individual branch operates only at a speed $F_S/L$, hence realization of a single ADC becomes quite feasible and economical at a nominal power expense. The digital outputs of each individual ADC $y_i[n]$ where  $i \in [1,L]$ are combined to result in the final digital output $y[n]$. \par Time-interleaved architectures, though seemingly very elegant, are plagued with issues attributed to mismatches between the individual branches. Three main mismatch sources can be identified, namely
\begin{figure}[t]
  \centering
  \includegraphics[width=0.35\textwidth]{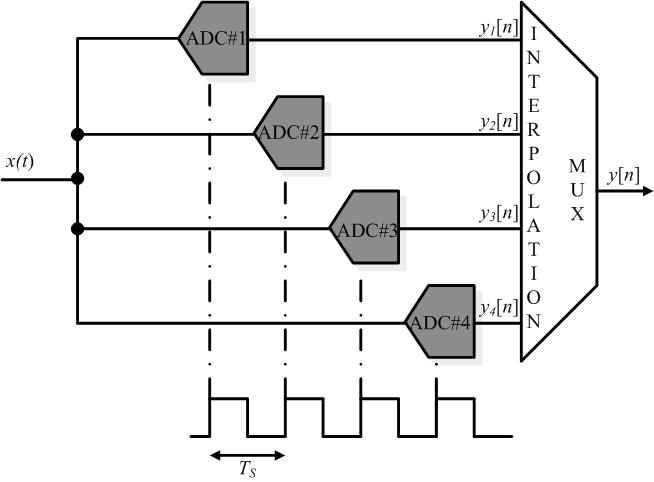}
  \caption{Time-Interleaved ADC: concept}
  \label{fig:simu}
\end{figure}

\begin{itemize}
\item DC offsets
\item Gain mismatches
\item Timing mismatches
\end{itemize}

\par Out of these three, the first two have been dealt in literature extensively, and satisfactory, low-cost solutions have been identified to mitigate their effects. The main ideas behind these solutions stem from estimating the aforementioned errors and then calibrating the system to counter them. The first two are easy to counter since channel-specific gain control and DC offset control/cancellation are easy to achieve in a power-economical way \cite{Eklund,Ndjountche,Ferragina}. However, calibrating for phase mismatches (timing mismatches) proves to be a much more tedious task \cite{Vogel1,Kurosawa}. Several techniques have been proposed in literature to alleviate the problem of timing-mismatches.

\subsection{Prior-art}
\par Most of the techniques to tackle timing mismatches are based on calibrating the inter-channel timing error(s) and then \textit{shifting} the sampling edge for each channel based on some form of digital filtering to match the correct edge using any form of digitally controlled delay elements \cite{Camarrero,Louwsma,ElChammas,Jamal,Vogel2}. Subtle differences can be found though, among these techniques and hence merit some discussion for developing the proposed idea. In \cite{Camarrero}, a digital detection subsystem (DDSS) and a digitally-controlled delay element (DCDE) operate in a negative feedback loop for each channel to adjust the timing edge based on the digital control word corresponding to each channel mismatch. The precision of the edge-matching obtained in this technique, however is based on the size of the digital-to-time converter step-size $t_{step}$ (which may be poorly controlled over process,voltage, temperature variations) and hence can limit the overall SNR for high performance systems. In \cite{Louwsma}, a master clock is used to synchronize all the sub-clocks, that thrusts a great reliance on the accuracy of laying out the channels symmetrically with a robust clock-distribution tree, which may not be feasible for most practical considerations. In \cite{ElChammas}, an interesting correlation based technique is used to estimate the timing-skews for each channel with respect to a reference channel. The estimation method relies on maximizing the SNR for each channel by moving the timing-edge (similar to gradient algorithms predominant in adaptive filters). The timing-edge is adjusted using a cascaded delay-line, implemented using variable capacitive loading through digital bits from the correlation mechanism. Although this work achieves excellent timing resolution but its dependence on PVT operating conditions makes it infeasible for a robust system. An interesting idea to correct for timing mismatches is proposed in \cite{Jamal}, where the ADC outputs are digitally filtered (to implement an all-pass transfer function) to impart phase delays corresponding to $\Delta{t}_i$ (phase skew in $i$-th channel). Practical considerations enforce the filters to be windowed using optimization algorithms. This technique, though somewhat independent of the \textit{minimum timing resolution}, will naturally operate better for very high-order filters rendering a large power consumption operating at the sub-ADC frequency.
\par In this paper, we propose a digital signal conditioning technique to allay the timing-mismatch errors. The main difference from most of the prior-art lies in the fact that an exact alignment of the sampling edge with the correct edge for a particular channel is \textit{not effected at all times}. However, it is ensured that on an average the sampling edge coincides with the exact edge. Also, the instantaneous errors, so committed are spectrally shaped out-of-band or scrambled depending on the resolution/power tradeoff.
\par It should be noted that the former (shaping the instantaneous errors) will be beneficial only for narrow signal bands of interest that are spread over a wide spectrum: particular examples of which maybe bandpass delta-sigma ADCs. The shaping technique in particular, consequently, is not applicable to Nyquist ADCs since out-of-band errors are not suppressed, thereby degrading the (signal-to-noise and distortion ratio)SNDR of the system. The proposed approach enables about 30dB of (spurious-free dynamic range)SFDR improvement over the uncorrected case, as detailed in Section III, decoupling the efficacy of the technique with the minimum timing resolution attainable in a particular CMOS technology node.
\par The paper is organized as follows. Section II presents the proposed technique. Behavioral simulation results are discussed in Section III while the paper is concluded in Section IV.

\section{Proposed Technique}
\subsection{Theory}
\par Let the input to the system be $x(t)$. The sampling time for the $i-th$ ADC, with a timing skew of $\tau_i$, is
\begin{align}
 t_{i}[n]=t_{i,ideal}[n]+\tau_{i}
\end{align}
where $t_{i,ideal}[k]$ is the ideal sampling time at the $k$-th time instant.
\begin{figure*}
  \centering
  \subfloat[DSC concept]{\label{fig:Xamplingphase}\includegraphics[width=0.27\textwidth]{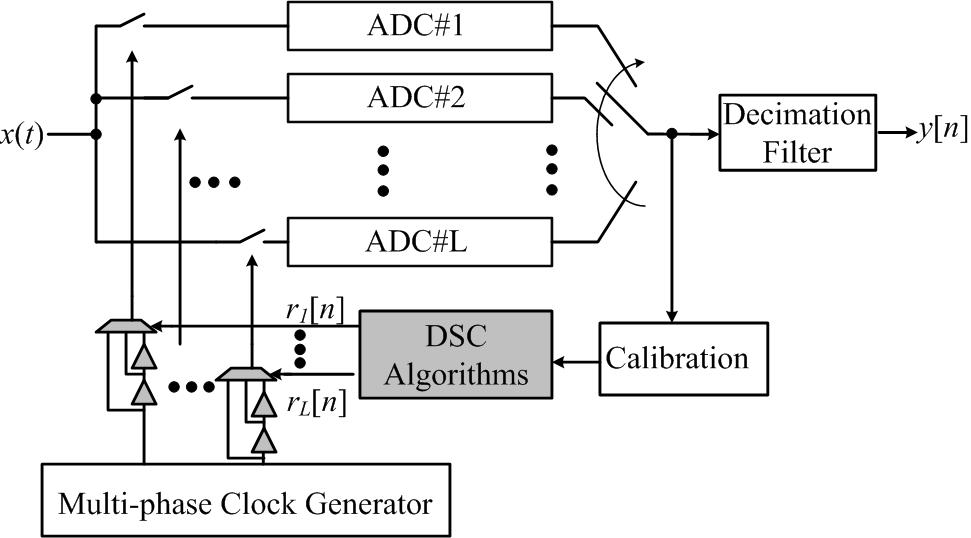}}
  \subfloat[DSC algorithm]{\label{fig:Xamplingphase}\includegraphics[width=0.27\textwidth]{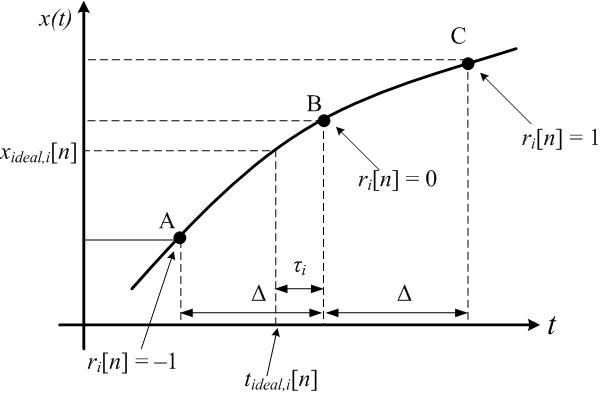}}
  \caption{Digital Signal Conditioning}
  \label{fig:simu}
\end{figure*}
\begin{figure*}
  \centering
  \subfloat[Scrambling]{\label{fig:Xamplingphase}\includegraphics[width=0.34\textwidth]{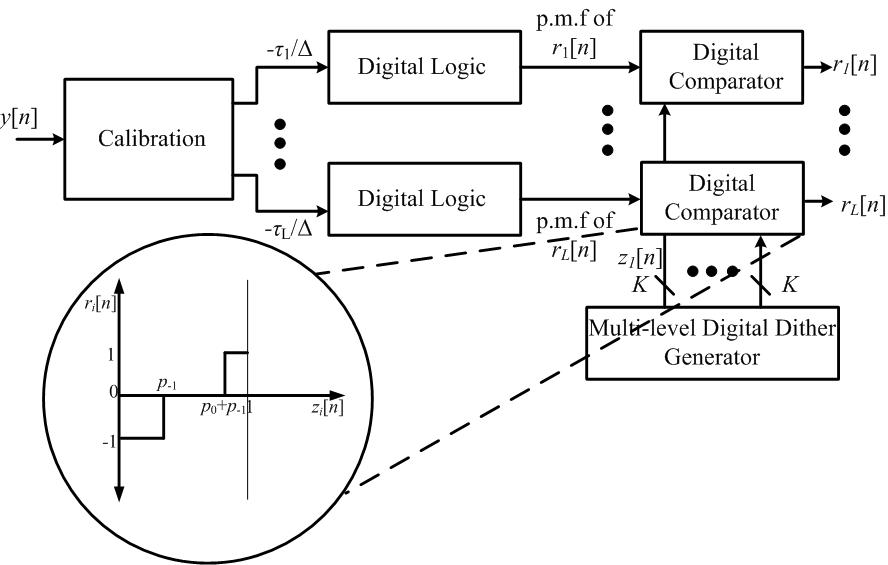}}
  \subfloat[Shaping]{\label{fig:Xamplingphase}\includegraphics[width=0.34\textwidth]{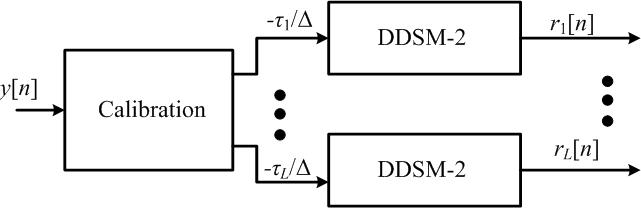}}
  \caption{DSC techniques}
  \label{fig:simu}
\end{figure*}

\par The proposed technique bears analogy with a fractional-N PLL multi-modulus divider \cite{Pamarti1}. Let $\Delta$ be a well-defined circuit quantity( $\Delta >> \tau_i$). Then it suffices to jump between the edges defined by $\mathcal{E}={t_i[n] \pm k\Delta},  \forall k \in \mathbb{N} \cup{\{0\}}$ in a definite manner to converge on to the exact edge in an average sense.
\par The proposed idea is illustrated in Fig. 2. With an estimate of the timing errors, $\tau_{i}$ between the branches (through post-calibration), a digital sequence $r_i[n]$ is generated for the $i$-th channel as shown. The different timing edges $\in \mathcal{E}_1 \subset \mathcal{E}$ are then chosen based on the sequence $r_i[n]$ (due to hardware constraints, a smaller set $\mathcal{E}_1$ is chosen). The cardinality of the set(and hence the span of $r_i[n]$) $\mathcal{E}_1$is determined based on the available hardware complexity and power-cost. Now, the sampling instant can be expressed as,
\begin{align}
 t_{i}[n]=t_{i,ideal}[n]+\tau_{i}+r_i[n]\Delta
\end{align}
\par Hence, the sampled signal for the $i$-th channel can be represented as (in a Taylor's series approximation)
\begin{align}
x_i[n] &=x(t)+(\tau_i+r_i[n]\Delta)\frac{dx(t)}{dt} \nonumber \\
       & +\frac{(\tau_i+r_i[n]\Delta)^2}{2}\frac{d^2x(t)}{dt^2}+......|_{t=t_{i,ideal}[n]}
\end{align}
\par Hence, it suffices to cancel the error terms $(\tau_i+r_i[n]\Delta)\frac{dx(t)}{dt}, \frac{1}{2}(\tau_i+r_i[n]\Delta)^2\frac{d^2x(t)}{dt^2}$ by an appropriate selection of the sequence $r_i[n]$. In other words, the signal conditioning techniques would attempt to generate sequences $r_i[n]$ that will ensure the following two conditions:
\begin{enumerate}[(a)]
\item $\mathbb{E}(\tau_i+r_i[n]\Delta)=0$
\item $\mathbb{E}((\tau_i+r_i[n]\Delta)^2)$ is bounded
\end{enumerate}
Based on this theory, we propose the scrambling technique.
\subsection{Scrambling the timing errors}

\par We choose the edge set $\mathcal{E}_1 ={t_i[n]-\Delta, t_i[n], t_i[n]+\Delta}$. We propose to select each one of these edges corresponding to the set $r_i[n]=\{-1,0,1\}$ with an element-wise correspondence. Let us assign the occurrence probabilities of the values of $r_i[n]$ as $p_{-1},p_0,p_1$ where the notations are self-explanantory. Then, we can construct,

\begin{align}
\left[\begin{array}{ccc} 1 & 1 & 1  \\
(1-\alpha_{i}) & -\alpha_{i} & -(1+\alpha_{i})  \\
(1-\alpha_{i})^2 & \alpha_{i}^2 & (1+\alpha_{i})^2 \end{array} \right]
\left[\begin{array} {c} p_{-1} \\ p_0 \\ p_1 \end{array} \right]
=\left[\begin{array} {c} 1 \\ 0 \\ g^2 \end{array} \right]
\end{align}
\par where $g$ is a small constant independent of all the ADCs and $\alpha_{i}=\tau_{i}/\Delta$ subject to the condition $\{p_{-1},p_0,p_1\} \in [0,1]$.
\par The three equations in Eqn. 4 can be understood intuitively:
\begin{itemize}
\item \{-1,0,1\} is the span of $r_i[n]$
\item The timing errors go to zero on an average
\item The second-order error term is a constant independent of the ADCs
\end{itemize}

\par \textit{Remark}: The last statement should be noted carefully. The error introduced due to the second term is independent of the individual ADCs and loses potency to cause spurious tones (for a tonal input). The error power is instead spread over the entire band (scrambling) \cite{Galton1}. Depending on the value of $g$ and the oversampling ratio in the ADCs, the resultant SNR hit needs to be evaluated. The higher-order terms may still engender some residual non-linearity, but they are negligible for all practical purposes.
\par Implementation of the digital signal conditioning (DSC) block in Fig. 2 is shown in Fig. 3(a). Eqn. 4 is solved to determine $\{p_{-1},p_0,p_1\}$. Subsequently, a $K$-bit digital dither ( $ \in [0,1]$) is passed through a quantizer with its thresholds set by $\{p_{-1},p_{-1}+p_0,1\}$ to generate the sequence $r_i[n]$. The technique is able to cut down the spurious tones by almost 20dB as illustrated in the next section.

\subsection{Spectrally shaping the timing errors}
\par The main idea proposed in this section is to generate the sequence $r_i[n]$ using a digital delta-sigma modulator (DDSM) as explained below. We assume having the estimate of the timing mismatch error $\tau_{i}$ in $K$ bits such that the selected edge converges in mean to the correct edge. The instantaneous error, so introduced, is now spectrally shaped out-of-band. A subtle but important point should be noted here. Even though the residual timing error is shaped out-of-band, since this error is different across channels, some residual second-order non-linearity may show up as spurious tones, but it can be proved theoretically (and substantiated through simulations), this error contribution is typically much lower than the accepted quantization-error of the ADC, and hence does not cause an issue. The technique is illustrated in Fig. 3(b). The negative of the timing-error for each channel, $-\tau_i$, scaled by the quantity $\Delta$, is passed through a digital delta-sigma modulator (DDSM) of order $P$ with $M$ output levels and the output of this DDSM is the control sequence $r_i[n]$. It should, however, be kept in mind that a DDSM is prone to producing idle tones unless certain conditions are satisfied \cite{He}. It has been proved in literature \cite{He}, that the input to the DDSM should be bounded in $[-(M+1-2^P)\frac{a}{2},(M+1-2^P)\frac{a}{2}]$, where $a$ is the step-size of the quantizer to prevent its overload. To further reduce the effect of limit-cycle tones,a small random signal $d[n]$ having the statistics $\mathbb{P}r(d[n]=0)=\mathbb{P}r(d[n]=1)=0.5$ is added to the $K$ bit input (LSB dithering) \cite{Pamarti2}. The output sequence $r_i[n]$ has an average of $\frac{-\tau_i}{\Delta}$ while its quantization error power is spectrally second-order shaped out of band \cite{Galton2}. The second-order error term for the $i$-th channel from Eqn. (3) is the dominant residual error(since the first-order term is driven to zero by the design of the DDSM).The analysis becomes intuitive for a tonal input. Let $x(t)=A\sin(\omega_0t)$. Then, evaluating the RHS of Eqn. 3 and extending to the cumulative error, we find,
\begin{align}
 e[n] &=-\sum_{i=1}^{i=L}\frac{1}{2}\mathbb{E}(\tau_i^2+r_i[n]^2\Delta^2+2\tau_ir_i[n]\Delta)\nonumber \\
      &A\omega_0^2\sin(\omega_0t)|_{t=t_{i,ideal}[n]} \nonumber \\
      &=-\sum_{i=1}^{i=L}\frac{1}{2}\mathbb{E}(-\tau_i^2+r_i[n]^2\Delta^2)A\omega_0^2\sin(\omega_0t)|_{t=t_{i,ideal}[n]} \nonumber \\
      &=\sum_{i=1}^{i=L}\frac{1}{2}(\tau_i^2-S_i{\Delta^2})A\omega_0^2\sin(\omega_0t)|_{t=t_{i,ideal}[n]}
 \end{align}
where $S_i \triangleq \mathbb{E}(r_i[n]^2)$. In this analysis, it is assumed that the channels are time-offset from a reference channel by the error $\tau_i$. Treating any one channel($k$-th one) as the reference channel makes $\tau_k'=0$ where $\tau_i'=\tau_i-\tau_k$. This observation does not distract from the given analysis and can be easily accounted for.
\par Now, it should be seen from \cite{Pamarti2}, that in a properly designed DDSM, since the error samples are independent of the input samples, hence for a $P-th$ order DDSM, the output variance of any $r_i[n]$ can be written as,
\begin{align}
\mathbb{E}(r_i[n]^2) &= \frac{a^2}{12}f(P) + \frac{\tau_i^2}{\Delta^2}
\end{align}
where $a$ is the step-size of the quantizer in the DDSM, $f()$ is a bounded function of the loop-filter order.
\par This error variance, is thus bounded for any given $P$ as long as the no-overload condition for the DDSM quantizer is satisfied.
In fact, the resultant SFDR of the ADC can be written as
\begin{align}
\text{SFDR}=10\log_{10}(\frac{4}{{\omega_0^4\sum_{i=1}^{i=L}(S_i{\Delta^2}-\tau_i^2)^2}})
\end{align}

 \section{Simulation Results}

 \begin{figure}[t]
  \includegraphics[width=0.4\textwidth]{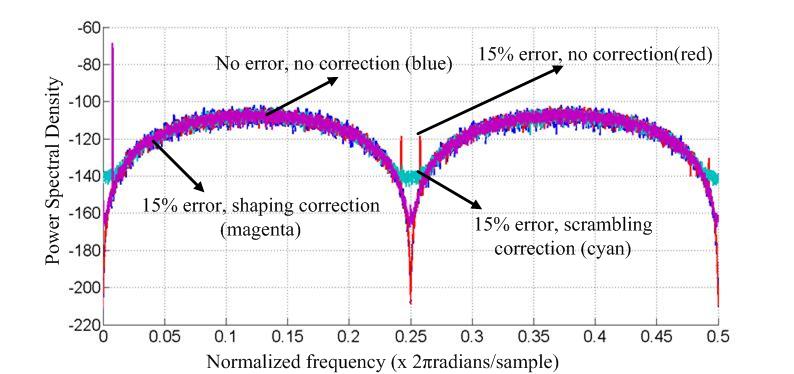}
 \caption{Output spectrum: different scenarios}
  \label{fig:simu}
\end{figure}
\par The simulation test-bench is described now. The input $x(t)$ is chosen to be a sinusoid. $L$, number of channels is chosen to be 4. Each channel ADC is configured to be a second-order sigma-delta ADC with 8 output levels. The shaping technique is shown for a second-order DDSM with four output levels i.e. $r_i[n] \in \{-3,-1,1,3\}$, which implies $a=2$ in Eqn. (6). For a 4-way interleaving, the noise-shaping nulls (from the actual ADC as well as from the DDSM) form at $F_{S}/4$. Timing-errors show up as spurious tones around the null at $F_{S}/4$. The ADC output spectrum is illustrated for the ideal scenario with no errors, for the timing errors with no correction as well as for both the techniques (scrambling and shaping) in Fig. 4. As can be seen, the scrambling technique eliminates almost all channel-dependent errors resulting in an elevated noise-floor mainly corresponding to the $g$ term from Eqn. 4. For the shaping case, the spectral nulls are much deeper as expected, the residual noise being attributed to the residual error terms in Eqn. 6. In fact, for a 2nd-order DDSM with four output levels, it can be shown that the SFDR due to 15\% timing error is about 90 dB, which should satisfy almost all practical communication scenarios. For the uncorrected case, the timing error begets spurious tones around the null at $\pi/2$, degrading the SFDR to about 60 dB.

\section{Conclusion}

\par Digital signal conditioning based techniques for mitigating timing mismatch errors in time-interleaved ADCs are proposed. Two power-efficient techniques, one based on scrambling the error components and another based on spectrally shaping the error components have been presented, their operations analyzed and their performances substantiated through behavioral simulations.

\end{document}